# Elimination of basal-plane stacking faults in semipolar/nonpolar GaN and light emitting diodes heteroepitaxially grown on sapphire substrates

Jie Song[1,2*], Joowon Choi[1,2], Cheng Zhang[1], Zhen Deng[1], Yujun Xie[3], and Jung Han[1]

[1]Department of Electrical Engineering, Yale University, New Haven, CT 06520, USA

[2]Saphlux Inc, Branford, CT 06405, USA

[3]Department of Mechanical Engineering & Materials Science, Yale University, New Haven, CT 06520, USA

*Email: jie.song@yale.edu

**ABSTRACT**

High quality semipolar and nonpolar GaN is crucial in achieving high-performance GaN-based optoelectronic devices, yet it has been very challenging to achieve large-area wafers that are free of basal-plane stacking faults (BSFs). In this work, we report an approach to prepare large-area, stacking-fault-free (SF-free) semipolar GaN on (4-inch) sapphire substrates. A root cause of the formation of BSFs is the emergence of N-polar $(000\bar{1})$ facets during semipolar and non-polar heteroepitaxy. Invoking the concept of kinetic Wulff plot, we succeeded in suppressing the occurrence of N-polar GaN $(000\bar{1})$ facets, and consequently in eliminating the stacking faults generated in $(000\bar{1})$ basal-planes. The result was confirmed by transmission electron microscopy, cathodoluminescence, and low-temperature photoluminescence characterizations. Furthermore, InGaN light emitting diodes with promising characteristics have been produced on the SF-free semipolar $(20\bar{2}1)$ GaN on sapphire substrates. Our work opens up a new insight about the heteroepitaxial growth of nonpolar/semipolar GaN and provides an approach of producing SF-free nonpolar/semipolar GaN material over large-area wafers which will create new opportunities in GaN optoelectronic and microelectronic research.

Semipolar and nonpolar crystallographic orientations of gallium nitride (GaN) have the possibility to address long-standing problems, such as efficiency droop and wavelength droop in III-Nitride light emitting diodes (LEDs) and laser diodes (LDs). A much-reduced polarization field in semipolar or nonpolar active regions is expected to greatly reduce the quantum confined Stark effect and improve carrier injection across heterointerfaces, thus leading to improved device performance in LEDs and LDs[1]. In the past decade, high output power blue and green LEDs and LDs have been demonstrated on semipolar planes such as ($10\bar{1}\bar{1}$), ($20\bar{2}1$), and ($30\bar{3}1$)[2–8]. However, all these high-performance devices could only be realized on bulk GaN substrates, which were produced by cross slicing of (0001)-oriented GaN crystals prepared by either hydride vapor phase epitaxy HVPE or ammonothermal growth[2–9]. The resulting nonpolar and semipolar substrates, while being of high microstructural quality with very low densities of stacking faults (SFs) and dislocations, are of very small and irregular dimensions defined by the shapes of as-grown crystals. The dimension and price of these substrates severely hinder the effort in mass-producing semipolar and nonpolar devices[2–9].

Since 2000, tremendous efforts were exerted on the heteroepitaxy of nonpolar/semipolar GaN on large-area, industry standard substrates including sapphire, silicon, or SiC substrates. These efforts evolved over time in approximate three generations. The first generation is planar heteroepitaxy of GaN directly on sapphire or SiC substrates. Planar heteroepitaxy of nonpolar and semipolar GaN suffered from a high density of SFs, including basal-plane SFs (BSFs) and prismatic SFs (PSFs) associated with BSFs[10–15]. BSFs and PSFs are very deleterious to the performance of GaN-based devices. Epitaxial lateral overgrowth (ELO) which was successful in suppressing dislocations on c-plane GaN was tested to reduce the density of SFs in semipolar and nonpolar GaN[16–20]. This second-generation attempt was found to be ineffective as lateral overgrowth along the N-polar ($000\bar{1}$) direction caused the formation of new SFs[16,17]. Several groups have developed an approach which we term orientation controlling epitaxy (OCE) to grow large-area semipolar GaN using patterned substrates[21–25]. This approach, designated as the third generation, was accomplished by selective area growth of GaN from inclined sidewalls on stripe-patterned sapphire or Si substrates. A schematic drawing about OCE of semipolar GaN on patterned sapphire or Si substrates is shown in Fig. S1 in Supplementary Information. During the OCE process, c-plane GaN is selectively grown on the exposed sapphire (0001) or Si (111) sidewalls. Once GaN grows out of trenches and coalesces with neighboring stripes to form a continuous film, a nonpolar or semipolar orientation of GaN is produced in the normal direction of the substrate. In spite of its flexibility in rendering almost arbitrary crystallographic orientations of GaN, the process of OCE still produces SFs albeit with a somewhat lower density[26–28]. It remains highly desirable to eliminate the SFs and produce SF-free semipolar or nonpolar GaN films on large-diameter substrates.

In this study, we report a method to achieve SF-free semipolar or nonpolar GaN by heteroepitaxial growth on large-sized foreign substrates. We will discuss the origin of the generation of SFs, disclosing the general principle of eliminating these otherwise persistent defects, and demonstrating a particular solution. As an example, we will demonstrate the elimination of SFs in semipolar $(20\bar{2}1)$ GaN and the growth of high crystalline quality 4-inch semipolar $(20\bar{2}1)$ GaN without SFs on sapphire substrates. The surface of the $(20\bar{2}1)$ GaN/sapphire template is micro-facetted with a combination of $(10\bar{1}0)$ and $(10\bar{1}1)$ facets, because of the thermodynamic properties of different GaN crystallographic planes. To demonstrate its utility, we performed chemo-mechanical polishing (CMP) to achieve atomically smooth semipolar GaN surface; high-performance InGaN LEDs are obtained on the SF-free semipolar $(20\bar{2}1)$ GaN/sapphire templates.

**Analysis of kinetic Wulff plot**

As reported previously, the presence of nitrogen-polar (N-polar) $(000\bar{1})$ facets interacting with foreign interface would typically result in the generation of SFs in semipolar or nonpolar GaN[15,27]. Schematic drawings about SFs generation in semipolar and nonpolar GaN are shown in Fig. S2 in the Supplementary Information. Similar results have also been observed by other groups in the ELO process of semipolar and nonpolar GaN that new SFs are always generated at the N-polar $(000\bar{1})$ face wing region[23,25,29]. Given the close correlation between the presence of N-polar basal $(000\bar{1})$ surface in heteroepitaxy and the occurrence of BSFs, it seems reasonable to assume that BSFs can be eliminated if the shape or morphology of growing GaN islands can be controlled such that the N-polar basal plane can be suppressed. The "shaping" of crystals can be achieved, through the concept of kinetic Wulff plot, by changing growth conditions[30,31] or even introducing additives in the form of surfactants[32]. In the specific case of GaN, it has been reported that impurities such as silicon (Si), magnesium (Mg), antimony (Sb), and Europium (Eu) can be used to alter the shape of growing GaN crystals[22,33–35].

To explore the influence of surfactants and doping on modifying the shape of GaN crystal, we have studied the kinetic Wulff plot by conducting selective area growth (SAG) of GaN with different dopants, including Si, germanium (Ge), and Mg, which are used as regular n-type and p-type doping, respectively. The SAG was conducted on SiO$_2$-masked openings of annular ring and circle pattern on a-plane GaN/sapphire templates. Schematic drawing about SAG is shown in Fig. S3 in the Supplementary Information. After growth, GaN crystal with different planes appears and the crystal planes can be identified and labeled in scanning electron microscopy (SEM) image, as shown in Fig. S3c and S3d. By studying the evolution of these facets, we can extract the growth rate of each plane and construct a kinetic Wulff plot. We first studied the GaN SAG with different doping and 45°-tilted-view SEM images of undoped, Si-doped, Mg-doped, and Ge-doped GaN SAG, respectively, can be found in Fig. S4 in the Supplementary Information. We find that the shape of GaN crystal and area of each facet varies with different dopants. One significant difference is that N-polar $(000\bar{1})$ facet disappears in Ge-doped GaN, while it appears in undoped, Si-doped and Mg-doped GaN. The evolution of GaN with different doping has been studied and the growth rate of each plane has been extracted according to the SAG results. 2-dimensional (2-D) projection kinetic Wulff plot, or more accurately the intersection of the 3-D contour with the plane encompassing c- and m-axes, is drawn for undoped and Ge-doped GaN, respectively, and orientations corresponding with stable GaN facets with the different color shown in Fig. 1a and 1b. More kinetic Wulff plots of GaN with Si-doping and Mg-doping can be found in Fig. S5 in the Supplementary Information. For undoped GaN as shown in Fig. 1a, N-polar $(000\bar{1})$ facet in undoped GaN is the facet with the slowest growth rate. However, as shown in Fig. 1b, the growth rate of N-polar $(000\bar{1})$ facet in Ge-doped GaN is dramatically enhanced and becomes much fast than $(10\bar{1}\bar{1})$. According to the theory of kinetic Wulff plot, in the convex growth mode, such as during the nucleation stage and selective growth before coalescence, the facet with fast growth rate would disappear and the crystal would be dominated by the facets with slow growth rates [30,31].

**OCE of semipolar $(20\bar{2}1)$ GaN**

As described earlier, OCE is a process of combining substrate patterning with selective area growth to "rotate" the surface orientation of GaN from conventional (0001) to nonpolar and semipolar directions. Semipolar (20$\bar{2}$1) GaN has been grown on patterned sapphire substrates (PSS) using OCE[28]. Fig. 1c shows a cross-sectional SEM image of OCE growth of GaN on a stripe patterned (22$\bar{4}$3) sapphire substrate (PSS). The inclined sidewall of PSS is c-plane sapphire and can support the growth of (0001) GaN at an inclined angle. As labeled in Fig. 1c, GaN stripes are bounded by three stable facets, including (10$\bar{1}$1), (10$\bar{1}$0) and (000$\bar{1}$), respectively. Upon further growth, the adjacent stripes coalesce to form a continuous GaN film with a surface normal direction along the [20$\bar{2}$1] direction (Fig. 2a). In spite of achieving the desired crystallographic orientation, cross-sectional transmission electron microscopy (TEM) images reveal the presence of a high density of SFs[27,28], typically up to a few times $10^5$ cm$^{-1}$.

Since Ge-doping is effective in shaping GaN to avoid the emergence of N-polar GaN facets, we tested Ge-doping during growth of semipolar (20$\bar{2}$1) GaN on PSS. Fig. 1d shows a cross-sectional SEM image of Ge-doped (20$\bar{2}$1) GaN grown on a patterned (22$\bar{4}$3) sapphire substrate. With a Ge doping of $2 \times 10^{19}$ cm$^{-3}$ (in reference to c-plane GaN doping), the growth rate of GaN in N-polar (000$\bar{1}$) direction towards left of SEM image is dramatically enhanced such that N-polar (000$\bar{1}$) facet no longer appears, and the GaN stripe crystal is now bounded by (10$\bar{1}$1), (10$\bar{1}$0), (10$\bar{1}\bar{1}$) and (1$\bar{1}$0$\bar{1}$) facets.

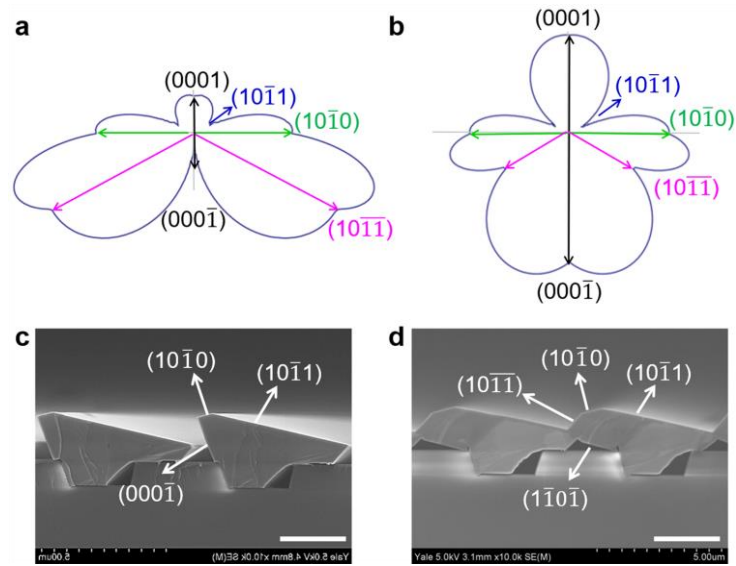

**Figure 1** 2-D kinetic Wulff plots mapped onto a-planes for (**a**) undoped and (**b**) Ge-doped GaN, respectively. Cross-sectional SEM images of (**c**) undoped and (**d**) Ge-doped (20$\bar{2}$1) GaN grown on patterned (22$\bar{4}$3)

sapphire substrate, respectively. The scale bars are 2 μm in both (**c**) and (**d**).

**Materials characterization results**

After the coalescence of adjacent Ge-doped GaN stripes, 8 μm un-doped GaN was subsequently grown to form a continuous GaN film with a total thickness of about 10 μm, as shown in Fig. 2a. The crystalline orientation of GaN is marked with surface normal direction towards GaN [20$\bar{2}$1]. An X-ray diffraction (XRD) 2θ/ω scan was conducted with a scan range from 30° to 90° to further confirm the crystal orientation, as shown in Fig. 2b. Only two diffraction peaks at 70.5° and 84.3° were observed, corresponding to GaN (20$\bar{2}$1) (interplanar spacing is 0.133 nm) and sapphire (22$\bar{4}$3) (interplanar spacing is 0.115 nm) diffractions, respectively, indicating that single-phase (20$\bar{2}$1) orientation GaN has been achieved with the direction parallel to the sapphire (22$\bar{4}$3) orientation. The crystalline quality of 10 μm thick GaN layers was also characterized by XRD rocking curves. On-axis XRD rocking curves of (20$\bar{2}$1) diffraction plane are shown in Fig. 2c, with the rocking axis in perpendicular (red curve) and parallel (blue curve) to patterned stripes, respectively. Full-width-at-half-maximums (FWHMs) of (20$\bar{2}$1) plane with the rocking axis perpendicular and parallel to patterned stripes are 192 and 217 arcsec, respectively. The FWHMs of (20$\bar{2}$1) GaN in this work are narrower than what we obtained before[27], indicating an improved quality of GaN is achieved in this work.

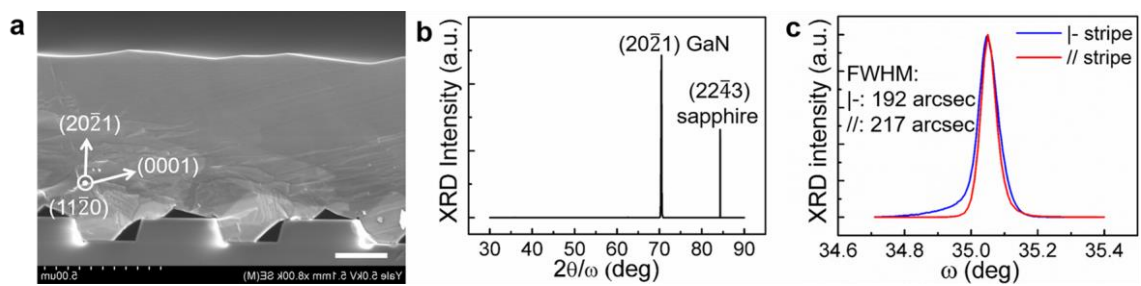

**Figure 2 a**, Thick undoped (20$\bar{2}$1) GaN regrown on Ge-doped GaN. **b**, XRD 2θ/ω scan showing single orientation GaN has been achieved, with the (20$\bar{2}$1) plane parallel to the sapphire substrate. **c**, XRD rocking curves of on-axis (20$\bar{2}$1) plane with rocking axis perpendicular (red curve) and parallel (blue curve) to patterned stripes, respectively. The scale bar is 2 μm.

The microstructural quality of OCE (20$\bar{2}$1) GaN grown on PSS was examined by transmission electron microscopy (TEM). Fig. 3a and 3b show bright-field cross-sectional TEM images of undoped and Ge-doping involved (20$\bar{2}$1) GaN, respectively, under two-beam condition. The top row (a1, b1) and bottom row (a2, b2) of TEM images were taken along diffraction vectors of $g$ = <0001> and $g$ = <10$\bar{1}$0>, respectively, with the zone axis of [11$\bar{2}$0]. The crystalline directions are labeled by the yellow arrows in Fig. 3. According to the extinction criteria of SFs in TEM image, SFs are visible with $g$ vector of <10$\bar{1}$0> and invisible with $g$ vector of <0001>[36]. Purely screw- and edge-types of threading dislocations (TDs) are revealed under diffraction vectors of $g$ = <0001> and $g$ = <10$\bar{1}$0>[37], respectively. As seen in Fig. 3a showing the TEM images of undoped (20$\bar{2}$1) GaN grown on PSS, besides the TDs generated at the interface between inclined c-plane sapphire and GaN, straight dark line contrast located in basal plane at N-polar GaN region is observed and attributed to the presence of BSFs[23,27], as marked by the red arrow in Fig. 3a2. However, for Ge-doping involved (20$\bar{2}$1) GaN, as shown in Fig. 3b1 and 3b2, only TDs are observed in the TEM images with diffraction vectors of both $g$ = <0001> and $g$ = <10$\bar{1}$0>. The occurrence and bending of TDs are in complete agreement with what has been reported in the epitaxial lateral overgrowth of c-plane GaN[38,39]. No dark line contrast in basal plane at N-polar GaN region is observed in Fig. 3b, indicating that no SFs have been generated in the Ge-doping involved (20$\bar{2}$1) GaN. This observation underscores the effectiveness of eliminating the basal-plane SFs in semipolar GaN by shaping the nucleation crystal to minimize the formation of N-polar basal plane of GaN in the lateral growth. Further plan-view TEM characterization has also confirmed that SF-free semipolar GaN has been achieved on sapphire (not shown here).

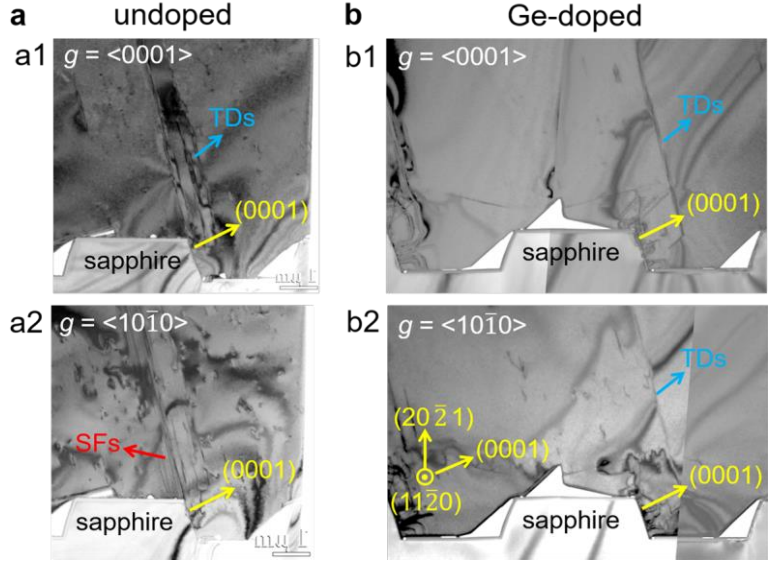

**Figure 3** Bright-field cross-sectional TEM images of (**a**) undoped and (**b**) involving Ge-doped (20$\bar{2}$1) GaN, respectively, under two-beam condition. The top row (**a1**, **b1**) and bottom row (**a2**, **b2**) were taken along diffraction vectors of ***g*** = <0001> and ***g*** = <10$\bar{1}$0>, respectively.

To further confirm the elimination of SFs, cathodoluminescence (CL) was conducted to characterize the (20$\bar{2}$1) GaN samples at room-temperature. Before CL measurement, samples were polished to minimize influence of surface morphology on CL signal since GaN (20$\bar{2}$1) facet is energetically unstable and consists of two stable facets, (10$\bar{1}$1) and (10$\bar{1}$0)[23,27]. Fig. 4a shows a plane-view SEM image of (20$\bar{2}$1) GaN after CMP, indicating that a smooth surface has been obtained. The surface of polished (20$\bar{2}$1) GaN was further examined by atomic force microscopy (AFM). Fig. 4c shows an AFM image with a scan area of 10 × 10 μm$^2$ and the root-mean-roughness (RMS) is around 0.5 nm. Plan-view panchromatic CL image is shown in Fig. 4b, taken from the same area shown in Fig. 4a. Defects including SFs and TDs would exhibit dark contrasts in CL scan image since these defects are non-radiative recombination centers. Usually, SFs as planar defects appear as dark line contrast in CL scan image[23,25,27]. A CL image of an undoped (20$\bar{2}$1) GaN grown on PSS is shown in Fig. S6 in Supplementary Information and dark line contrast is observed due to the existing of BSFs. As seen in Fig. 4b, there is no dark line contrast observed, indicating that there is no SF generated in our (20$\bar{2}$1) GaN. Only dark spots are observed, which are attributed to the presence of TDs with an estimated density of around 1 × 10$^8$ cm$^{-2}$. We have examined different positions on a 4-inch (20$\bar{2}$1) GaN by CL and showed that we could achieve SF-free semipolar (20$\bar{2}$1) GaN across the entire 4-inch wafer. Fig. 4d shows a picture of SF-free (20$\bar{2}$1) GaN/sapphire templates, including both 2-inch and 4-inch.

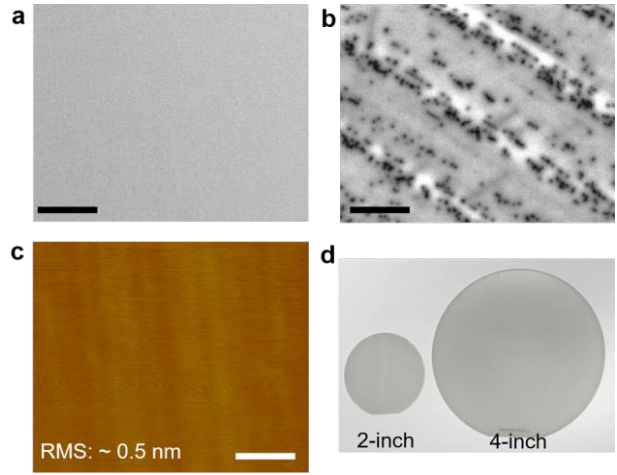

**Figure 4 a**, Plan-view SEM image of (20$\bar{2}$1) GaN after CMP. **b**, Plan-view panchromatic CL scan image of (20$\bar{2}$1) GaN. **c**, AFM image of (20$\bar{2}$1) GaN after CMP. **d**, A photo of a polished 4-inch SF-free (20$\bar{2}$1) GaN grown on the patterned sapphire substrate. The scale bars are 5 μm in both (**a**) and (**b**), and 2 μm in (**c**), respectively.

Low-temperature photoluminescence (LT-PL) was conducted to supplement the microstructural study of 10 μm Ge-doped GaN. As it has been reported that the presence of BSFs will give rise to distinct PL features at low temperature[40–42]. Two (20$\bar{2}$1) GaN samples on sapphire were studied. One is a standard, undoped OCE GaN template while the other is with Ge-doping. At 10K, near-band-edge (NBE) emission at 3.48 eV is observed for both undoped and Ge-doped (20$\bar{2}$1) GaN. As shown by the blue curve in Fig. 5, strong BSF related emission at 3.44 eV and PSFs related emission at 3.32 ~ 3.35 eV are observed in undoped (20$\bar{2}$1) GaN, indicating the presence of high density of BSFs and PSFs. However, for Ge-doped (20$\bar{2}$1) GaN, both BSFs and PSFs related emission disappear, indicating that both BSFs and PSFs have been significantly suppressed in Ge-doped (20$\bar{2}$1) GaN. The peak located at 3.41 eV observed in both undoped (as a shoulder near to BSF emission peak in blue curve) and Ge-doped (20$\bar{2}$1) GaN is likely attributed to the impurity-bounded excitons[43,44] since the O impurities are highly incorporated in semipolar (20$\bar{2}$1) GaN[45].

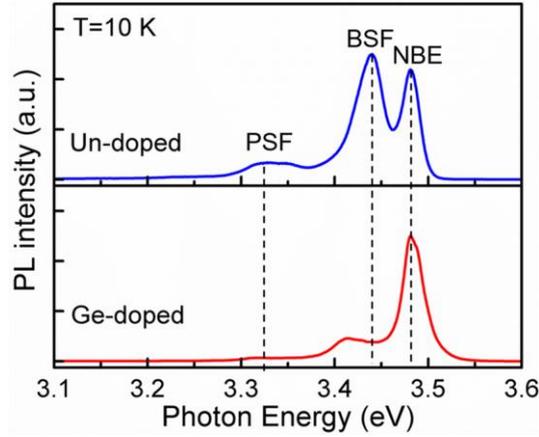

**Figure 5** LT-PL of thick GaN grown on undoped (blue curve) and Ge-doped (20$\bar{2}$1) (red curve) GaN on PSS, respectively.

## Semipolar (20$\bar{2}$1) LED results

InGaN-based blue LEDs were grown on CMP-polished SF-free (20$\bar{2}$1) GaN/sapphire template. A semipolar (20$\bar{2}$1) GaN bulk substrate with size of 5 × 10 mm$^2$ bought from Mitsubishi Chemical was co-loaded together with the SF-free (20$\bar{2}$1) GaN/sapphire template. Room-temperature electroluminescence (EL) measurements were performed on LEDs under continuous-wave current injection conditions. Fig. 6a shows an I-V curve of a LED with a turn-on voltage of about 3.2 V, which is comparable with a typical turn-on voltage of a c-plane LED. A low leakage current of ~ 10$^{-6}$ A has been achieved under a reverse-bias of -5 V. A photo of a (20$\bar{2}$1) LED operated at 40 mA is shown in the inset of Fig. 6a. EL spectra of LED grown on (20$\bar{2}$1) GaN/sapphire template are shown in Fig. 6b with an increasing injection level from 20 to 140 mA under continuous-wave current injection conditions at room temperature. The linewidth of the EL spectrum at an injection current of 100 mA is about 22 nm, similar to a typical c-plane LED (~ 21 nm). The light output power (LOP) of the LEDs grown on (20$\bar{2}$1) GaN/sapphire template and (20$\bar{2}$1) GaN bulk substrate is shown by the red and blue spots, respectively, in Fig. 6c. As shown in Fig. 6c, the performance of LED grown on (20$\bar{2}$1) GaN/sapphire template is about 70% of that on a (20$\bar{2}$1) GaN bulk substrate at the injection current of above 100 mA. Previously all the reports of semipolar and nonpolar LEDs grown on sapphire showed a performance of at least a factor of 10 or more reduction in output power, with less than 1 mW at an injection current of up to 200 mA[46–49]. The preliminary comparison of LEDs grown on the Ge-doped GaN PSS templates with the bulk substrates provides a clear evidence of the benefit from the elimination of SFs. The difference in performance could be explained by the presence of TDs, as shown by CL in Fig. 4b, with a concentrated density of about $4 \times 10^8$ cm$^{-2}$ along growth fronts aligned with stripe trench pattern. Further reduction of dislocations and possible misfit dislocations due to preexisting dislocations[50] should bring it closer to the mass production of semipolar GaN LEDs.

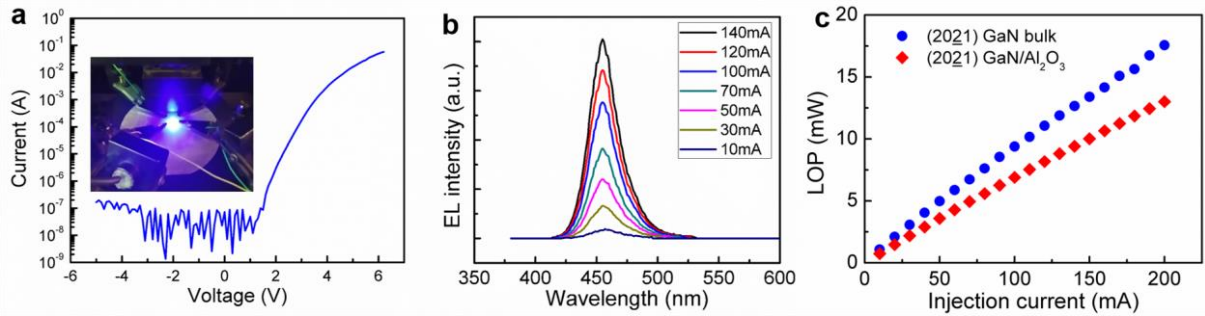

**Figure 6 a**, I-V curve of a LED die (Inset shows a photo of lighting (20$\bar{2}$1) LED). **b**, EL spectra with different injection currents under continuous-wave current injection operation. **c**, LOP of LEDs grown on (20$\bar{2}$1) GaN/sapphire template (red) and (20$\bar{2}$1) GaN bulk substrate (blue), respectively.

In summary, we have demonstrated producing large-area SF-free semipolar GaN on large-area sapphire substrates by introducing a Ge-doping during the growth to suppress the formation of N-polar (000$\bar{1}$) facets. We found that Ge-doping can accelerate the growth rate of N-polar (000$\bar{1}$) facet and slow down the growth rate of (10$\bar{1}\bar{1}$), resulting in the elimination of N-polar (000$\bar{1}$) facets and subsequent SFs generated in (000$\bar{1}$) facets. We have exhibited eliminating the SFs in semipolar (20$\bar{2}$1) GaN and growing 4-inch SF-free semipolar (20$\bar{2}$1) GaN on sapphire, confirmed by the characterizations of TEM, CL and LT-PL. This is the world-first demonstration about large-area SF-free semipolar GaN heteroepitaxially grown on a foreign substrate. Furthermore, InGaN LEDs have been grown on the SF-free semipolar (20$\bar{2}$1) GaN on sapphire substrates, with the performance very similar to that on GaN bulk substrate. Our work opens up a new insight about the heteroepitaxial growth of nonpolar/semipolar GaN and provides an approach of producing SF-free nonpolar/semipolar GaN material over large-area wafers which will create new opportunities in GaN optoelectronic and microelectronic research.

**Methods**

**MOCVD growth and device fabrication:** MOCVD growth was performed in a Aixtron low-pressure MOCVD reactor. Growth of a (20$\bar{2}$1) oriented GaN layer was conducted on PSS. The sapphire substrates were chosen such that the [0001] sapphire direction with respect to the substrate surface corresponds to exactly the [0001] GaN direction with respect to the (20$\bar{2}$1) GaN surface, i.e., c-plane offcut 75.09° towards the a-direction (sapphire (22$\bar{4}$3) offcut 0.45°). Linear trenches of nominally 3 μm wide with a pitch of 6 μm and a depth of 1 μm are prepared by photolithography with reactive-ion etching. Except for the desired c-plane sapphire sidewall, the entire sapphire surfaces are masked by dielectric $SiO_2$, fabricated with a self-aligned technique of angled evaporation. More details about stripe patterned sapphire substrates and schematic drawing about OCE process can be found in our previous publication[27,28]. After preparation, the sapphire substrates are loaded into the MOCVD reactor chamber for growing (20$\bar{2}$1) GaN.

After achieving (20$\bar{2}$1) GaN grown on sapphire, the wafer was taken out of the reactor and planarized by a CMP machine to reach a smooth surface in order to conduct InGaN LEDs growth on (20$\bar{2}$1) GaN/sapphire templates. LED structure consists of a 1.5-μm-thick Si-doped GaN with a Si-doping concentration of about $1 \times 10^{19}$ $cm^{-3}$, followed by three pairs of InGaN (4 nm)/GaN (6 nm) multiple quantum wells active region. The active region was followed by a 150 nm P-type GaN layer with a Mg-doping concentration of about $7 \times 10^{19}$ $cm^{-3}$. Subsequently, small-area (600 × 600 $μm^2$) mesas were processed by standard photolithography and Cl-based reactive ion etching. Ni (20 nm)/Au (50 nm) was used as both n- and p-type contacts.

**Materials and device Characterization:** The GaN samples were characterized by SEM, AFM, XRD, LT-PL and CL. The microstructural properties were analyzed by TEM. EL was conducted at room temperature to characterize the performance of semipolar (20$\bar{2}$1) InGaN LEDs.

**Acknowledgment**

We acknowledge research support from Saphlux Inc. and use of facilities supported by YINQE and NSF MRSEC DMR 1119826.

**Authors contribution**

J.S. and J.H. conceived the initial concept. J.S. and Z.D. conducted SAG experiments. J.S. and C.Z. analyzed and drew kinetic Wulff plots. J.S. performed MOCVD growth and materials characterization. Y.J.X. fabricated TEM specimen. J.W.C. grew LEDs and conducted LED device fabrication and EL measurement. J.S. and J.H. wrote the manuscript.

**Competing interests**

Jung Han is a co-founder of Saphlux and acknowledges that he has a significant financial interest with Saphlux.

**Materials & Correspondence**

Correspondence and requests for materials should be addressed to J.S.

# Supplementary Information for

# Elimination of basal-plane stacking faults in semipolar GaN and light emitting diodes heteroepitaxially grown on sapphire substrates

**Schematic drawing of OCE**

Fig. S1 shows the schematic drawing about process of orientation control epitaxy (OCE) for semi/non-polar GaN grown on patterned sapphire substrate (PSS) or Si substrates. **a**, Choose the right sapphire or Si substrates. **b**, Pattern the substrate into stripe trenches with one sidewall nearly parallel to sapphire c-plane or Si(111) plane. **c**, Deposit dielectric mask on the surface of patterned substrate, except the sidewalls which are parallel to sapphire c-plane or Si(111) plane. **d**, Selectively grow GaN on exposed sidewalls of patterned substrate. **e**, Continue to grow GaN and coalese the stipes to form a thick continues semipolar GaN film. The normal surface direction will be GaN semipolar direction.

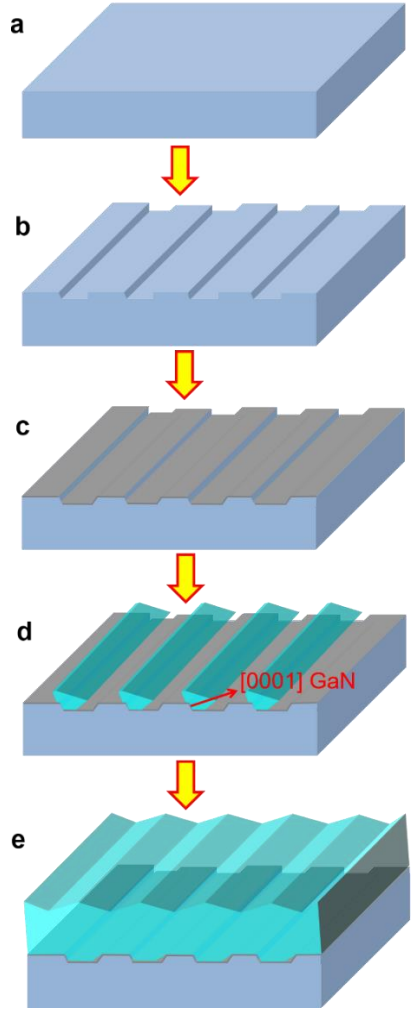

**Figure S1** Schematic drawing about semipolar GaN grown on patterned sapphire substrates. **a**, Choose the right sapphire or Si substrates. **b**, Pattern the substrate into stripe trenches with one sidewall nearly parallel to sapphire c-plane or Si(111) plane. **c**, Deposit dielectric mask on the surface of patterned substrate, except the sidewalls which are parallel to sapphire c-plane or Si(111) plane. **d**, Selectively grow GaN on exposed sidewalls of patterned substrate. **e**, Continue to grow GaN and coalese the stipes to form a thick continues semipolar GaN film. The normal surface direction will be GaN semipolar direction.

**Schematic drawing about the generation of stacking-faults**

Fig. S2a and S2b show the schematic drawing about the growth of semipolar and nonpolar GaN, respectively, on patterned sapphire substrates. The GaN facets have been labeled in Fig. S2. Once the N-polar $(000\bar{1})$ basal-plane GaN appears on the surface of dielectric mask, the basal-plane stacking faults will be generated in $(000\bar{1})$ GaN above the dielectric mask, as shown by the black lines in Fig. S2a and S2b.

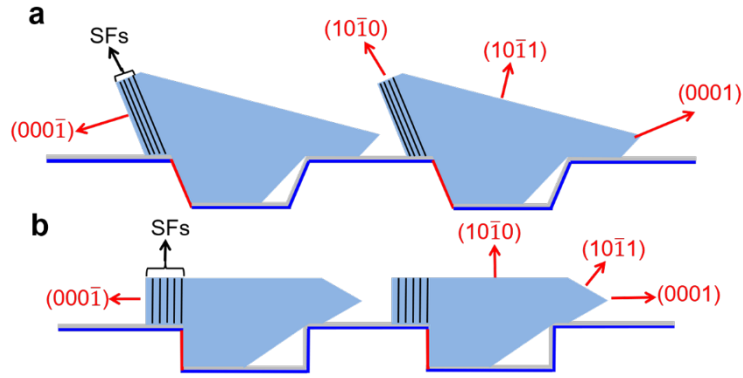

**Figure S2 a**, Schematic drawing about semipolar GaN grown on patterned sapphire substrates. **b**, Schematic drawing about nonpolar GaN grown on patterned sapphire substrates. Basal-plane stacking faults are generated in (000$\bar{1}$) GaN, as shown by black lines in the figure.

**Study of SAG and kinetic Wulff plot**

The selective area growth (SAG) was conducted on SiO$_2$-masked openings of annular ring and circle pattern on a-plane GaN/sapphire templates, as shown in Fig. S3a and S3b. After growth, GaN crystal with different planes appears and the crystal planes can be identified and labeled in scanning electron microscopy (SEM) image, as shown in Fig. S3c and S3d. Figure S3c and S3d show the top-view and 45°-tilted-view SEM images of GaN SAG on an annular ring pattern, respectively.

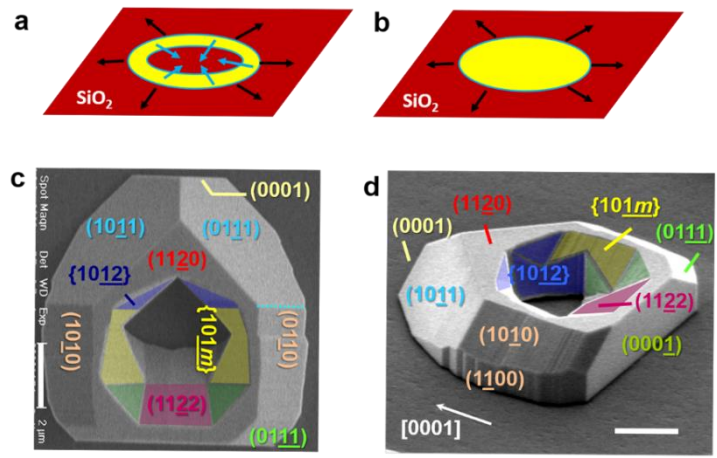

**Figure S3** Schematic drawing of (**a**) annular ring and (**b**) circle pattern, respectively. (**c**) Top-view and (**d**) 45°-tilted-view SEM images of GaN SAG on an annular ring pattern. The scale bars are 2 μm in both (**c**) and (**d**).

Fig. S4a and S4b show 45°-tilted-view SEM images of undoped, Si-doped ([Si] = 1 E19 cm$^{-3}$), Mg-doped ([Mg] = 5E19 cm$^{-3}$), and Ge-doped ([Ge] = 2E19 cm$^{-3}$) GaN SAG, respectively. The top and bottom rows in Fig. S4 show GaN grown on annular ring and circle opening, respectively. The shape of GaN crystal and area of each facet varies with different dopants. One significant difference is that N-polar (000$\bar{1}$) facet disappears in Ge-doped GaN, while it appears in undoped, Si-doped and Mg-doped GaN.

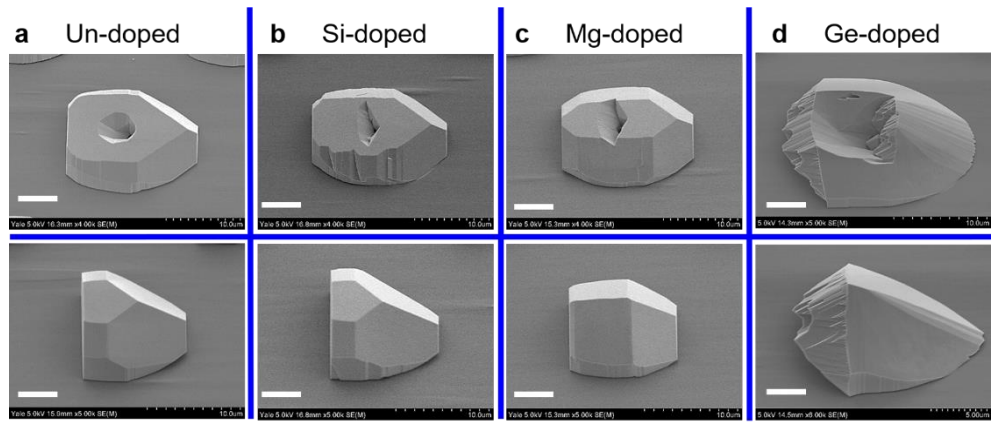

**Figure S4** 45°-tilted-view SEM images of (**a**) undoped, (**b**) Si-doped, (**c**) Mg-doped and (**d**) Ge-doped GaN SAG, respectively. The top and bottom rows show GaN grown on annular ring and circle opening, respectively. The scale bars are 2 μm.

2-dimensional (2-D) kinetic Wulff plots mapped onto a-planes have been drawn for GaN grown with different types of doping, respectively, and GaN facets have been marked with different color shown in Fig. S5, including un-doping, Si-doping, Mg-doping and Ge-doping. The growth rates of different facets can be varied dramatically by different dopants. N-polar (000$\bar{1}$) facet is the facet with the slowest growth rate in undoped, Si-doped and Mg-doped GaN. However, the growth rate of N-polar (000$\bar{1}$) facet in Ge-doped GaN is dramatically enhanced and becomes much fast than (10$\bar{1}\bar{1}$).

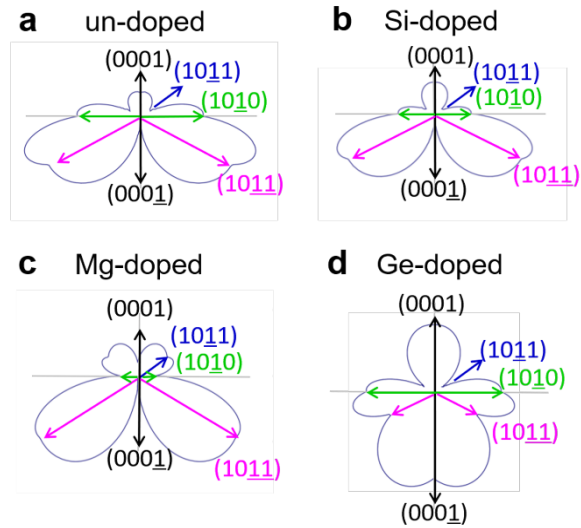

**Figure S5** 2-D kinetic Wulff plots mapped onto a-planes for GaN grown with (**a**) un-doping, (**b**) Si-doping, (**c**) Mg-doping and (**d**) Ge-doping, respectively. GaN facets have been marked with different color.

## CL characterization of undoped (20$\bar{2}$1) GaN grown on PSS

Fig. S6 shows a panchromatic CL image of an undoped (20$\bar{2}$1) GaN grown on PSS. The dark band observed on the image are caused by the concentrated basal-plane stacking faults. The dark spots are corresponding to the threading dislocations.

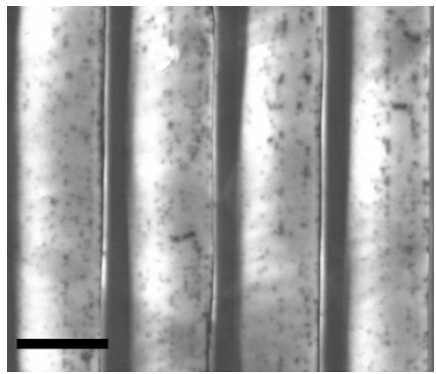

**Figure S6** Panchromatic CL image of an undoped (20$\bar{2}$1) GaN grown on PSS. Scale bar is 5 μm.